\documentclass[useAMS]{mn2e}
\usepackage{color}
\usepackage{graphicx}

\title{Long-term X-ray emission from Swift J1644+57}
\author[Y. C. Zou, F. Y. Wang and K. S. Cheng] {Y. C. Zou$^{1,2}$\thanks{zouyc@hust.edu.cn}, F. Y. Wang$^{1,3,4}$\thanks{fayinwang@nju.edu.cn}
and K. S. Cheng$^1$\thanks{hrspksc@hku.hk}\\
{$^1$Department of Physics, The University of Hong Kong, Pokfulam
Road, Hong Kong, China}
\\
{$^2$School of Physics, The Huazhong University of Science and Technology, Wuhan 430074, China} \\
{$^3$School of Astronomy and Space Science, Nanjing University,
Nanjing 210093, China}\\
{$^4$Key Laboratory of Modern Astronomy and Astrophysics (Nanjing
University), Ministry of Education, Nanjing 210093, China}}

\begin{document}
\maketitle

\begin{abstract}
The X-ray emission from Swift J1644+57 is not steadily decreasing
instead it shows multiple pulses with declining amplitudes. We model
the pulses as reverse shocks from collisions between the late
ejected shells  and the externally shocked material, which is
decelerated while sweeping the ambient medium. The peak of each
pulse is taken as the maximum emission of each reverse shock. With a
proper set of parameters, the envelope of peaks in the light curve
as well as the spectrum can be modelled nicely.
\end{abstract}


\section{Introduction}
Tidal disruption of a star by a supermassive black hole in a
galactic nucleus has been investigated by many authors (Hills 1975,
Lacy et al. 1982). When a star's trajectory happens to be
sufficiently close to a supermassive black hole, the star will be
captured and eventually tidally disrupted. After the star is
disrupted, at least half of the debris is ejected from the system,
the remainder remains bound to the black hole and is accreted (Rees
1988; Ayal et al. 2000). The accretion of this stellar debris has
been predicted to power a luminous electromagnetic flare that is
expected to peak in the optical, ultraviolet(UV) and X-ray
wavelengths, lasting for months to years (Ulmer 1999; Stubbe \&
Quataert 2009, 2011).

Swift J164449.3+573451 (also known as GRB 110328A in the beginning,
hereafter Swift J1644+57) has been proposed as a tidal disruption
candidate (Bloom et al. 2011; Levan et al. 2011; Burrows et al.
2011; Zauderer et al. 2011; Reis et al. 2012). Swift J1644+57 was
initially discovered as a long-duration gamma-ray burst (GRB
110328A) by the Swift Burst Alert Telescope (BAT). Swift follow-up
observations with the Ultraviolet and Optical Telescope (UVOT) and
X-ray Telescope (XRT) began 1475 s after the initial trigger. No
source was seen in the UVOT observations, but a bright point source
was found with the XRT (Bloom et al. 2011). It remained bright and
highly variable for a long period, and re-triggered the BAT three
times. This high-energy transient is unlike any known events, such
as active galactic nuclei (AGN), or gamma-ray bursts (GRBs). The BAT
and XRT spectral fits are consistent during the brightest flaring
stage, and well fitted with a broken power-law model (Burrows et al.
2011). Considering the repeated extremely short timescale X-ray
flares, Wang \& Cheng (2012) proposed a later internal shock model
for the X-ray flares of Swift J1644+57, in which the reverse shock
is relativistic and the forward shock is Newtonian. From the strong
emission lines of hydrogen and oxygen, the redshift of Swift
J1644+57 is $z \sim 0.35$ (Levan et al. 2011). From the X-ray,
optical, infrared, and radio observations, it is found that the
position of this source is consistent with the nucleus of the host
galaxy (Bloom et al. 2011). This event has not been detected by the
Fermi Large Area Telescope (LAT) or VERITAS (Aliu et al. 2011). Only
upper limits were given in the GeV and TeV bands.

The peak isotropic luminosity of Swift J1644+57 is about $\sim
2\times 10^{48} \rm~erg~s^{-1}$ (Bloom et al. 2011; Burrows et al.
2011), which is about $10^4$ times larger than the Eddington
luminosity of a $10^6M_\odot$ black hole. Due to this
super-Eddington luminosity and the position of Swift J1644+57
relative to the center of its host galaxy, it is suggested that
Swift 1644+57 is powered by a relativistic jet created by accretion
onto a $10^{6-7}M_\odot$ black hole (Bloom et al. 2011; Burrows et
al. 2011; Shao et al. 2011). The jet Lorentz factor was limited to
$\Gamma\leq 20$ from high-energy observations (Burrows et al. 2011),
the event rate (Burrows et al. 2011), and radio observations
(Zauderer et al. 2011; Berger et al. 2012). Wong et al. (2007) first
argued that the X-ray emission of a tidal disruption event may not
necessarily come from radiation of the accretion disk alone.
Instead, it may be related to a jet. As the jet travels in the
interstellar medium, a shock is produced and synchrotron radiation
is expected. They compared the light curve and the synchrotron
radiation spectrum in their model with the observed data and found
that the model can well explain the observed transient X-ray
emission from NGC 5905 and the late-time spectrum. Some theoretical
models have been proposed to explain unusual features of this event,
e.g. Swift J1644+57 could be explained as a white dwarf disrupted by
a $10^4M_\odot$ black hole (Krolik \& Piran 2011) or a tidal
obliteration event, which disrupts a star in a deeply plunging orbit
at periastron (Cannizzo et al. 2011). Lei \& Zhang (2011) argued
that the jet is launched by the Blandford-Znajek mechanism
(Blandford \& Znajek 1977) and found the spin parameter of the
central black hole is very high. Further, Lei, Zhang \& Gao (2013)
inferred the inclination angle between the black hole spin axis and
the star orbit. Saxton et al. (2012) studied the X-ray timing and
spectral evolution of this event and found that the spectrum became
mildly harder in its long-term evolution. Gao (2012) argued that the
outflow should be Poynting-flux dominated. De Colle et al (2012)
have shown that the stochastic contribution of the luminosity due to
the feeding rate variability induced by instabilities can explain
the X-ray light curve of Swift J1644+57 using a two-dimensional
simulation. Berger et al. (2012) and Metzger et al. (2012) modelled
the radio emission of Swift J1644+57 using the GRB afterglow model.
Berger et al. (2012) also found that the energy increase cannot be
explained with continuous injection from an $L\propto t^{-5/3}$ tail
and the relativistic jet has a wide range of Lorentz factors. Most
recently, Zauderer et al. (2013) found that the X-ray flux has a
sharp decline by a factor of 170 at about 500 days.

Based on the late X-ray light curve, which is composed of multi
X-ray flares, we propose that the X-ray emission comes from the
reverse shock produced by the late ejected shell colliding with the
decelerating material, which is decelerated by the ambient medium
through which it moves. The whole scenario of our model is as
follows: at the beginning, the central engine ejects relativistic
shells with higher velocity (corresponding to higher Lorentz
factor). They catch up with the outermost slower shell and the
consequent reverse shocks produce the early ($1-10^5$ s) X-rays, as
has been proposed by Wang \& Cheng (2012). At later times, the
central engine ejects shells with lower velocity (lower Lorentz
factor), and at the same stage, the outermost shell has been
decelerated by the medium, which \textbf{drives} an external shock.
This external shock produces the radio emission (Metzger et al.
2012; Berger et al. 2012). Then ejected shells collide with this
outermost decelerating material, and the reverse shocks produce the
late X-rays, which decreases with time as the density of the
emitting region decreases. The late injection of the shells also
enhances the total energy of the external shock, which provides the
needed energy resource as to explain the late enhancement of the
radio emission presented by Berger et al. (2012). In this paper, we
mainly focus on the later X-rays emission, which is produced by
collisions between ejected shells and the decelerating material. We
describe our model in section \ref{model}, and conclude in section
\ref{conclusion}.

\section{Modelling} \label{model}
The X-ray emission shows very rapidly flaring in the whole radiation
period from the beginning until several $10^7$ seconds. This
suggests that the flares should not come from the continuous
external shock. However, the external shock does exist, and may emit
mainly at optical and radio frequencies when the early ejected
shells combine and sweep the ambient medium\footnote{ As the
emission of the external shock very sensitively relies on the
Lorentz factor (see Zou \& Piran 2010 for more details), for the
relatively low Lorentz factor of this tidally disruption event
compared with that of a normal GRB, the flux density and the typical
frequencies will be much lower than in GRBs. }. With the continuous
activity of the central engine fed by material still
\textbf{remaining} around the central black hole, the late ejected
shells will eventually catch up with the external decelerating
shock, and a reverse shock emerge back into the ejected shells.
Berger et al. (2012) also found that the outflow is structured and
the relativistic jet was produced with a wide range of Lorentz
factors. Consequently, the forward shock will proceed into the
shocked medium. However, if the number density of the shocked medium
is high enough (corresponding to a weak non-relativistic forward
shock, which depends on the contrast of the density between the two
regions (Sari \& Piran 1995) ), this forward shock will be weak
enough and the corresponding emission would be negligible; and the
bulk Lorentz factor of the reverse shock will be the same as that of
the external shocked material. If the energy of the shell is much
smaller than the kinetic energy of the external shock, the influence
on the dynamics of the external shock can be neglected. The
behaviour of the external shock can be described in the same way as
a standard GRB afterglow (Sari, Piran \& Narayan 1998).

The scenario for the late X-ray emission in our model is as follows:
the continuous external shock mainly emits radio and optical
emission, and the late ejected shells collide on the decelerating
external shocked material, the reverse shock of each collision
produces one flare of the X-ray emission. Superposition of all the
flares by these episodically ejected shells composes the whole late
flaring X-ray light curve. The following are the details of the
scaling laws:

The Lorentz factor of the external shock is (Sari, Piran \& Narayan
1998)
\begin{equation}
\gamma \simeq 4.3 \,n_{0}^ {- {{1}\over{8}}
}\,(1+z)^{{{3}\over{8}}}\,E_{k,0,54
 }^{{{1}\over{8}}}\,t_{\oplus,6}^ {- {{3}\over{8}} },
\end{equation}
where $n$ is the number density of the medium, $z$ is the redshift,
$E_{k,0}$ is the isotropic kinetic energy of the external shock, and
$t_{\oplus}$ is the observer's time. The notation $Q=10^x Q_x$ is
used throughout the paper. The evolution of the Lorentz factor is
consistent with the radio observation at 216 days after the BAT
trigger (Berger et al. 2012). For example, at $t_{\oplus}=216$ days,
the Lorentz factor is about 2.2 for $E_{k,0}=5\times 10^{53}$ erg
and $n=0.2$ cm$^{-3}$ from observation. At the late time, the
Lorentz factor $\gamma$ has decreased into a relatively low region,
and the edge of the jet will be seen by the observer for a normal
jet opening angle $\theta_j=0.1$. AGN jets are also confirmed from
super-massive black holes. The apparent opening angle can be as high
as tens of degrees in an AGN jet, while the intrinsic opening angle
could be as low as a few degrees (Pushkarev et al. 2009). So our
choice of $\theta_j \sim 0.1$ is reasonable. For relativistic
motion, the relation between radius and observing time is ${\rm d}r
= 2 (1+z) \gamma^2 c {\rm d} t_{\oplus}$. The radius evolution with
time is then
\begin{equation}
  r \simeq 2.71 \times 10^{18}\,n_{0}^ {- {{1}\over{4}} }\,(1+z)^ {- {{1
 }\over{4}} }\,E_{k,0,54}^{{{1}\over{4}}}\,t_{\oplus,6}^{{{1}\over{4
 }}} ~{\rm cm}.
 \end{equation}

We denote the reversely shocked and unshocked regions of the ejected
shell as regions 3 and 4 respectively, and the forwardly shocked and
unshocked medium as regions 2 and 1 respectively. The average
Lorentz factor of protons in the reverse shock (region 3) is
$\bar\gamma_3 \simeq
\frac{1}{2}(\frac{\gamma_4}{\gamma_3}+\frac{\gamma_3}{\gamma_4})
\simeq \frac{\gamma_4}{2\gamma_3}$ (if $\gamma_4 \gg \gamma_3$),
where $\gamma_3=\gamma$ is the bulk Lorentz factor of the reverse
shock, and $\gamma_4$ is the bulk Lorentz factor of the ejected
shell. We get
\begin{equation}
\bar\gamma_3 \simeq
14.88\,n_{0}^{{{1}\over{8}}}\,\gamma_{4,2}\,(1+z)^ {- {{3}\over{8}}
}
 \,E_{k,0,54}^ {- {{1}\over{8}} }\,t_{\oplus,6}^{{{3}\over{8}}}.
\end{equation}

Because of the spread of velocities inside a single shell, at times
later than $10^6$s the spreading effect dominates the width of the
ejected shell, which is $r/2 \gamma_4^2$ (Sari \& Piran 1995). The
duration of the reverse shock is determined by the competition of
the width of the shell and the jet angular size, i.e.,
$\max(\frac{r}{2 \gamma_4^2 c}, \frac{r \theta_j^2}{2 c}) = \frac{r
\theta_j^2}{2 c}$, as $\gamma_4 \sim 100$, $\theta_j \sim 0.1$ in
this case. Then, the duration of each pulse is
\begin{equation}
\delta T \simeq 4.5 \times 10^5 \,n_{0}^ {- {{1}\over{4}}
}\,(1+z)^{{{3}\over{4}}}\,
 \theta_{j,-1}^2\,E_{k,0,54}^{{{1}\over{4}}}\,t_{\oplus,6}^{{{1
 }\over{4}}}~ {\rm s}.
 \end{equation}
The value of the duration is several times larger than the observed
duration of the individual pulses. However, notice $\delta T$ is
very sensitive to the jet opening angle. It requires the value of
$\theta_j$ to be a bit smaller than $0.1$.

Similar to the treatment in Zou, Wu \& Dai (2005), the number
density of the reverse shock region is
\begin{eqnarray}
n_{3} &=& (4 \bar\gamma_3 +3) n_4 \nonumber \\ & \simeq & 1.8 \times
10^{-2} \,n_{0}^{{{5}\over{8}}}\,\gamma_{4,2}^ {- 1 }\,(1+z)^{{{9
 }\over{8}}}\,E_{k,0,54}^ {- {{5}\over{8}} }\,E_{k,4,53}\,
 t_{\oplus,6}^ {- {{9}\over{8}} }~ {\rm cm^{-3}},
\end{eqnarray}
where $n_4$ is the number density of the ejected shell, which is
determined by the Lorentz factor $\gamma_4$ and the total kinetic
energy $E_{k,4}$. The internal energy density of the reverse shock
is
\begin{eqnarray}
e_{3} &=& \bar \gamma_3 n_3 m_p c^2 \nonumber \\ & \simeq & 4.0
\times 10^{-4}\,n_{0}^{{{3}\over{4}}}\,(1+z)^{{{3}\over{4}}
 }\,E_{k,0,54}^ {- {{3}\over{4}} }\,E_{k,4,53}\,t_{\oplus,6}^ {- {{3
 }\over{4}} }~ {\rm erg\,cm^{-3}}.
\end{eqnarray}

After the dynamical values are settled on, we follow the method to
get the synchrotron radiation of Sari, Piran \& Narayan (1998). The
electrons are accelerated into a power-law distribution:
$N(\gamma_e) d\gamma_e=N_\gamma \gamma_e^{-p}d\gamma_e
(\gamma_e>\gamma_m)$, where $\gamma_e$ is the randomized electron
Lorentz factor, $\gamma_m$ is the minimum Lorentz factor of the
accelerated electrons and $p$ is the power-law index. Assuming that
constant fractions $\epsilon_e$ and $\epsilon_B$ of the internal
energy go into the electrons and the magnetic field, we have the
magnetic field $B_3=\sqrt{8\pi\epsilon_B e_3}$, where $e_3$ is the
internal energy density of the shocked material. One  gets $\gamma_m
= \epsilon_e (\bar{\gamma_3}-1)(m_p/m_e)(p-2)/(p-1)$, and
$N_\gamma=n_3 (p-1) \gamma_m^{p-1}$. The peak spectral power of the
synchrotron emission for one electron is
\begin{eqnarray}
P_{\nu,\max} &=& (1+z) \sigma_T m_e c^2 \gamma B_3/ (3 q_e) \nonumber \\
 &\simeq& 4.0 \times 10^{-23}\,n_{0}^{{{1}\over{4}}}\,(1+z)^{{{7}\over{4
 }}}\,E_{k,0,54}^ {- {{1}\over{4}} }\,E_{k,4,53}^{{{1}\over{2}}}\,
 \nonumber \\ & & \varepsilon_{B,-1}^{{{1}\over{2}}}\,t_{\oplus,6}^ {- {{3}\over{4}} }
 ~{\, \rm erg \, Hz^{-1} \, s^{-1} } ,
\end{eqnarray}
where $\sigma_T$ is the Thomson cross section, and $q_e$ is the
electron charge. The peak observed flux density is then
\begin{eqnarray}
f_{\nu,\max} &=& (\gamma \theta_j)^2 {N_e P_{\nu,\max} }/{(4\pi D^2)} \nonumber \\
&\simeq& 4.8 \times 10^{-27}\,\gamma_{4,2}^ {- 1 }\,D_{28}^ {- 2
}\,(1+z)^{{{5}\over{2}}}\,\theta_{j,-1}^2\,E_{k,4,53}^{{{3}\over{2}}}\,
\nonumber \\ & & \varepsilon_{B,-1}^{{{1}\over{2}}}\,t_{\oplus,6}^
{- {{3}\over{2}} }~ {\rm erg\, cm^{-2} \, Hz^{-1} \, s^{-1}},
\end{eqnarray}
where $N_e=E_{k,4}/(\gamma_4 m_p c^2)$ is the total isotropic
equivalent number of electrons in region 3,  and $D$ is the
luminosity distance. Here we need to consider the beaming factor
$(\gamma \theta_j)^2$ $\simeq 0.11 \,n_{0}^ {- {{1}\over{4}}
}\,(1+z)^{{{3}\over{4}}}\,\theta
 _{j,-1}^2\,E_{k,0,54}^{{{1}\over{4}}}\,t_{\oplus,6}^ {- {{3}\over{4
 }} }$, as the jet opening angle is smaller then $1/\gamma$, and the isotropic
solution should be corrected by the beaming factor (Rhoads 1999),
while the lateral expansion is not considered for simplicity, a fact
also supported by numerical simulations (Cannizzo et al. 2004). With
$z=0.35$, the corresponding luminosity distance is $D\simeq 5.7
\times 10^{27}$cm in a cosmological model $\Omega_M=0.27$ and
$\Omega_\Lambda=0.73$ (Wright 2006).

The cooling Lorentz factor $\gamma_c$ is defined such that the
electron with $\gamma_c$ approximately radiate all its kinetic
energy in the dynamical time, i.e., $(\gamma_c-1)m_e
c^2=P(\gamma_c)t_{co}$, where $P(\gamma_e)=(4/3)\sigma_T
c(\gamma_e^2-1)(B^2/8\pi)$ (Rybicki \& Lightman 1979) is the
synchrotron radiation power of an electron with Lorentz factor
$\gamma_e$ in the magnetic field $B$, and $t_{co}$ is the dynamical
time in the comoving frame. Then the cooling Lorentz factor is
$\gamma_c \simeq {6\pi m_e c}{\sigma_T B^2 t_{co}}$, where $\sigma_T
\simeq 6.65\times 10^{-25} {\rm cm}^2$ is the Thompson scattering
cross-section.

The typical frequency (in the observer's frame) for a given electron
is $\nu_{\rm syn} = 3(1+z)^{-1}\gamma \gamma_e^2 q_e B/(2\pi m_e
c)$. The critical frequencies of the synchrotron emission are
\begin{eqnarray}
\nu_m &=& 3(1+z)^{-1}\gamma \gamma_m^2 q_e B/(2\pi m_e c)\nonumber \\
&\simeq& 2.5 \times
10^{12}\,n_{0}^{{{1}\over{2}}}\,\gamma_{4,2}^2\,\varepsilon_{e,-{{1}\over{2}}}^2\,(1+z)^
{- 1 } \,E_{k,0,54}^ {- {{1}\over{2}}}\,E_{k,4,53}^{{{1}\over{2}}}\,
\nonumber \\ & & \varepsilon_{B,-1}^{{{1}\over{2}}}~ {\rm Hz}, \\
\nu_c &=& 3(1+z)^{-1}\gamma \gamma_c^2 q_e B/(2\pi m_e c)\nonumber\\
&\simeq& 7.7 \times 10^{16}\,n_{0}^ {- {{1}\over{2}} }\,(1+z)^ {- 2
}\, \theta_{j,-1}^ {- 4 }\,E_{k,0,54}^{{{1}\over{2}}}\,E_{k,4,53}
 ^ {- {{3}\over{2}} }\,\nonumber \\ & &\varepsilon_{B,-1}^ {- {{3
 }\over{2}} }\,t_{\oplus,6}~ {\rm Hz}.
 \end{eqnarray}
The synchrotron self absorbing frequency $\nu_a$ for
$\nu_a<\nu_m<\nu_c$ is (Zou, Wu \& Dai 2005)
\begin{eqnarray}
 \nu_a &=& 1.1 \times 10^{8}\,n_{0}^{{{1}\over{5}}}\,\gamma_{4,2}^ {- {{8}\over{5}} }\,\varepsilon_{e,-{{1
 }\over{2}}}^ {- 1 }\,(1+z)^{{{1}\over{5}}}\,E_{k,0,54}^ {- {{1
 }\over{5}} }\,E_{k,4,53}^{{{4}\over{5}}}\,\nonumber \\ & &\varepsilon_{B,-1}^{{{1
 }\over{5}}}\,t_{\oplus,6}^ {- {{6}\over{5}} }~ {\rm Hz}.
 \end{eqnarray}

The formulae for the flux density in different spectral segments
were shown in Sari, Piran \& Narayan (1998). The flux density
of the late external reverse shock region $f_{\nu}(t_\oplus)$ (in units of ${\rm erg\,cm^{-2}\,Hz^{-1}\,s^{-1}}$) for different segments are:\\
for $\nu < \nu_{a} < \nu_{m} < \nu_{c} :$
\begin{eqnarray}
f_{\nu} \simeq 1.37 \times 10^{-10}\,n_{0}^ {- {{1}\over{2}}
}\,\gamma_{4,2}\,\nu_{17}^2\,D_{28}^ {- 2 }\,\varepsilon_{e,
 -{{1}\over{2}}}  
 \,(1+z)^{{{5}\over{2}}}\,\theta_{j,-1}^2\,E_{k,0,
 54}^{{{1}\over{2}}}\,t_{\oplus,6}^{{{1}\over{2}}},
 \label{f1}
\end{eqnarray}
$\nu_{a} < \nu < \nu_{m} < \nu_{c} :$
\begin{eqnarray}
f_{\nu} &\simeq & 1.65 \times 10^{-25}\,n_{0}^ {- {{1}\over{6}}
}\,\gamma_{4,2}^ {- {{5}\over{3}} }\,\nu
 _{17}^{{{1}\over{3}}}\,D_{28}^ {- 2 }\,\varepsilon_{e,-{{1}\over{2}}
 }^ {- {{2}\over{3}} } 
 \,(1+z)^{{{17}\over{6}}}\,\theta_{j,-1}^2\,E
 _{k,0,54}^{{{1}\over{6}}}\,\nonumber \\ & &E_{k,4,53}^{{{4}\over{3}}}\,\varepsilon_{
 B,-1}^{{{1}\over{3}}}\,t_{\oplus,6}^ {- {{3}\over{2}} },
 \label{f2}
\end{eqnarray}
$\nu_{a} < \nu_{m} < \nu < \nu_{c} :$
\begin{eqnarray}
f_{\nu} &\simeq & 8.28 \times
10^{-30}\,n_{0}^{{{3}\over{10}}}\,\gamma_{4,2}^{{{1}\over{5}}}\,\nu_{17}^
{- {{3
 }\over{5}} }\,D_{28}^ {- 2 }\,\varepsilon_{e,-{{1}\over{2}}}^{{{6
 }\over{5}}}  
 \,(1+z)^{{{19}\over{10}}}\,\theta_{j,-1}^2\,E_{k,0,54}
 ^ {- {{3}\over{10}} }\,\nonumber \\ & &E_{k,4,53}^{{{9}\over{5}}}\,\varepsilon_{B,-1
 }^{{{4}\over{5}}}\,t_{\oplus,6}^ {- {{3}\over{2}} },
 \label{f3}
\end{eqnarray}
$\nu_{a} < \nu_{m} < \nu_{c} < \nu :$
\begin{eqnarray}
f_{\nu} &\simeq & 7.26 \times
10^{-30}\,n_{0}^{{{1}\over{20}}}\,\gamma_{4,2}^{{{1}\over{5}}}\,\nu_{17}^
{- {{11
 }\over{10}} }\,D_{28}^ {- 2 }\,\varepsilon_{e,-{{1}\over{2}}}^{{{6
 }\over{5}}} 
 \,(1+z)^{{{9}\over{10}}}\,E_{k,0,54}^ {- {{1}\over{20}} }
 \,\nonumber \\ & &E_{k,4,53}^{{{21}\over{20}}}\,\varepsilon_{B,-1}^{
 {{1}\over{20}}}\,t_{\oplus,6}^ {- 1 }.
 \label{f4}
\end{eqnarray}

The X-rays observed by XRT are in the range of 0.3keV--10keV,
corresponding to $\nu_1 = 7.2 \times 10^{16}$ Hz--$\nu_2= 2.4 \times
10^{18}$ Hz. Equations (\ref{f3}) and (\ref{f4}) are corresponding
to the X-rays. As $\nu_m$ does not change with time, to fit the
slope of the X-ray light curves (the \textbf{trend} of the peaks),
it indicates that the $\nu_c$ was crossing the observed band.

\begin{figure}
 \includegraphics[width=0.5\textwidth,angle=0]{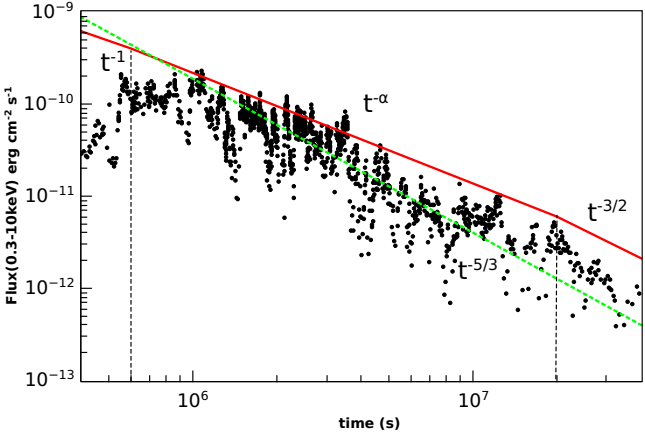}
 \caption{The X-ray light curve of Swift J1644+57.
Data are taken from http://www.swift.ac.uk/xrt\_curves/00450158/
(Evans et al. 2009). The three segments line is the envelope for the
peaks from the model, while the vertical dashed lines indicates the
time for the change of time behaviour because of the frequency
crossing. $\alpha$ is varying with time from 1 to 1.5. The dotted
line is a $t^{-5/3}$ slope shown for comparison.}
 \label{fig-flux}
\end{figure}

With a suitable tuning to model the observational light curves and
the spectral evolution, we get a set of proper parameters which are
far from extreme: $p=1.8$\footnote {Notice that as $p=1.8<2$, a
cut-off Lorentz factor of the shock electrons $\gamma_{\max}$ should
be introduced to make the model self-consistent, and it may change
the $\gamma_m$ a \textbf{little} bit (Dai \& Cheng 2001). However,
the value of $\gamma_{\max}$ depends on different models (Dai \&
Cheng 2001, Bhattacharya 2001), and the introduction of
$\gamma_{\max}$ into $\gamma_m$ will make the expressions much more
complicated. To keep the simplicity, we choose not to take this
$\gamma_{\max}$ effect into account.}, $n=1 {\rm cm^{-3}}$,
$E_{k,4}=3\times 10^{53}$ erg, $E_{k,0}=3\times 10^{54}$ erg,
$\varepsilon_B = 0.1$, $\varepsilon_e=0.3$, $\gamma_4=70$,
$\theta_j=0.05$. These parameters including the Lorentz factor of
the external shock $\gamma$ are consistent with the constraint from
the radio observations. Berger et al. (2012) found that the Lorentz
factor of external shock is $\gamma\sim 2.2-6.0$, the ambient
density is $0.2-60$ cm$^{-3}$, the jet energy is larger than
$5\times10^{53}$ erg using the radio observation up to 216 days
after the BAT trigger. Metzger et al. (2012) found that
$\epsilon_e=0.03-0.1$, the ambient density is $1-10$ cm$^{-3}$, and
the opening angle is 0.01-0.1.

\begin{figure}
 \includegraphics[width=0.5\textwidth]{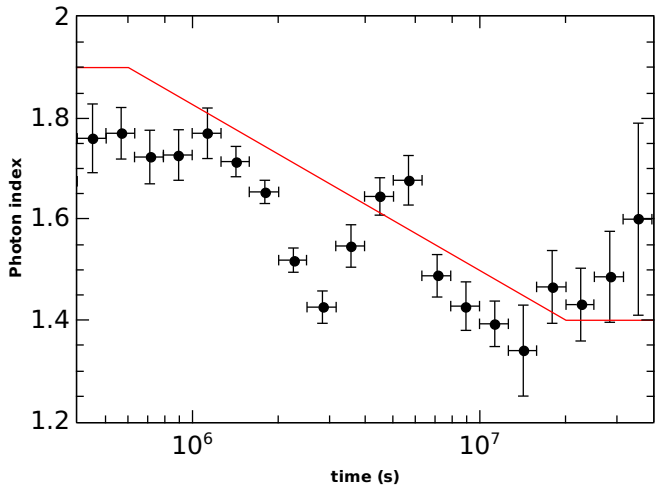}
 \caption{Photon index  of Swift J1644+57 versus time.
Data are taken from http://www.swift.ac.uk/xrt\_spectra/00450158/.
The solid line is the index from the model for the peak times, which
roughly dominates the whole spectrum.}
 \label{fig-spectra}
\end{figure}

The scaling laws with time of the derived quantities are: Lorentz
factor of the emitting region $\gamma \simeq 4.3 \, t_{\oplus,6}^ {-
{{3}\over{8}} }$, which decreases with time, but may be slower than
$t_{\oplus,6}^ {- {{3}\over{8}} }$ as new injected shells speed up
the external shock a little bit, which depends on the total energy
of the late ejection. The duration of each pulse is $\delta T \simeq
1.9 \times 10^5 \, t_{\oplus,6}^{{{1}\over{4}}}$ s, which is
consistent with the duration of the pulse in the observed light
curve. The slight widening with time is also reported by Saxton et
al. (2011). The characteristic frequency $\nu_{m} \simeq 8.1 \times
10^{11}$ Hz, which does not change with time and is always below the
X-ray frequencies of XRT. The cooling frequency $\nu_c \simeq 1.3
\times 10^{17}\, t_{\oplus,6}$ Hz, divides the X-ray band in several
epochs, for $t_{\oplus} < 6 \times 10^5$ s,
$\nu_m<\nu_c<\nu_1<\nu_2$, with X-ray spectral index $-\frac{p}{2}$,
i.e., -0.9, corresponding to the photon index 1.9. This value is
somewhat bigger than a rough average of the observed photon index
1.8. The reason is that $\nu_c$ is calculated for the time when the
reverse shock just crosses the shell ($\nu_c$ decreases from the
beginning until this time). Therefore, for each pulse, there is a
period when $\nu_c$ is higher than the observed frequency and the
case is $\nu_m < \nu < \nu_c$, and then the average photon index
could be smaller than 1.9. For the time $6 \times 10^5$ s
$<t_{\oplus} < 2 \times 10^7$ s, $\nu_m<\nu_1<\nu_c<\nu_2$, with
X-ray spectral index between $(-\frac{p}{2},-\frac{p-1}{2})$, i.e.,
(-0.9,-0.4), and for $t_{\oplus} > 2 \times 10^7$ s,
$\nu_m<\nu_1<\nu_2<\nu_c$, with X-ray spectral index
$-\frac{p-1}{2}$, i.e., -0.4. All the corresponding photon indices
are shown in Fig. \ref{fig-spectra}. The flux densities of
synchrotron radiation  in different ranges are: $f_\nu \simeq 6.1
\times 10^{-28} t_{\oplus,6}^{-1}~ {\rm erg \, cm^{-2} \, Hz^{-1} \,
s^{-1}}$ for $\nu_m<\nu_c<\nu_1<\nu_2$, which is suitable for time
$t< 6 \times 10^5$s. At $t = 6 \times 10^5$s, $f_\nu \simeq 1.0
\times 10^{-27}$, and the flux is $4.0 \times 10^{-10}~ {\rm erg \,
cm^{-2} \, s^{-1}}$. This is well consistent with the observation as
seen in Fig. \ref{fig-flux}. $f_\nu \simeq 5.3 \times 10^{-28}
t_{\oplus,6}^{-\frac{3}{2}} {\rm erg \, cm^{-2} \, Hz^{-1} \,
s^{-1}}$ for $\nu_m<\nu_1<\nu_2<\nu_c$, which is suitable for time
$t > 2 \times 10^7$ s. In the meanwhile between ($ 6 \times 10^5$ s,
$2 \times 10^7$ s), the case is $\nu_m<\nu_1<\nu_c<\nu_2$, and the
flux should be contributed to by both bands. The temporal index
should be in between $-1$ and $-\frac{3}{2}$, and the photon index
should be between 1.4 and 1.9 during this period, which is shown in
Fig. \ref{fig-spectra}. We can clearly see the transition of the
photon index from this figure, which can be well understood by the
model (solid line). Although during the transitional period ($ 6
\times 10^5$ s, $2 \times 10^7$ s) the photon indices varying with
time does not trace the solid line, this might come from the
diversity of the late ejected shells. For times earlier than $ 6
\times 10^5$ s, the predicted flux is higher than the observed one.
This might be caused by the transition of the external shock from a
coasting phase into a decelerating phase.

\section{Conclusion and discussion} \label{conclusion}
In this paper, we have proposed a scenario for the late X-ray
emission from Swift J1644+57, namely, a central engine produces
long-lasting episodic relativistic ejecta. These ejecta catch up
with the decelerating material, which sweeps the medium, and the
reverse shock propagating back into the ejecta emit the observed
flaring X-rays. As the ejecta are shells but not a steady jet, the
emission appears as flares. As shown in Figs. 1 and 2, our model is
well consistent with the observed peaks of the flares on both the
envelope of the light curve and the photon indices.

This scenario inherits the `prompt' emission picture, as proposed by
Wang \& Cheng (2012). Both of them are from the reversed shock of
the central engine ejection, which gives a complete scenario for
this event. At the beginning, the central engine produces shells
with higher velocity. They catch up the outermost slower shell and
the consequent reverse shocks produce the early (1-$10^5$ s) X-rays.
At later times, the central engine eject shells with relatively
lower velocity (assumed to be with the same Lorentz factor
$\gamma_4$ for the whole late times), and at the same stage, the
outermost shell has been decelerated by the medium, which
\textbf{drives} an external shock. Then ejected shells collide with
this decelerating material, and the reverse shocks produce the late
X-rays, which decreases with time. The late injection of the shells
also enhance the total energy of the external shock, which provides
the needed energy resource to explain the late enhancement of the
radio emission. Our model predicts the X-ray peaks will continue its
$t^{-1.5}$ slope until the external shock dominates the radiation,
which will appear shallower with a less variable light curve. This
can be examined by the follow-up observations.

Although the scenario for the late X-rays is a follow up to the
scenario for the earlier X-ray emission proposed by  Wang \& Cheng
(2012), there are some significant differences. First, the objective
is different. Wang \& Cheng (2012) treated the early emission; while
this paper treats the late X-ray emission with obvious flare.
Secondly, the scenario is different. The reverse shock is produced
by collision between later ejected shells and decelerating material
in this paper. But the reverse shock in Wang \& Cheng (2012) is
produced by two shells colliding. The physical origin of the reverse
shock is quite different. Thirdly, in this paper, the width of the
reversely shocked region 3 is much longer than that of Wang \& Cheng
(2012), because of the spreading of shells. So the duration of the
reverse shock is also long. The duration of X-ray pulse is about
$10^5$ s from observation (Saxton et al. 2011), which is consistent
with our model (see Eq. 4). In our model, as the relativistic jet
shuts off, there is no shell catching up with the decelerating
material, the emission of the reverse shock (mainly in X-rays)
disappears. The X-ray emission of the external shock is weak, which
is consistent with Swift and Chandra observations (Zauderer et al.
2013).

The origin of the hard electron distribution ($p<2$) is not yet
clear. The observations also confirmed the evidence for hard
electron distribution in some GRBs (Dai \& Cheng 2001; Huang et al.
2006; Covino et al. 2010). Simulations of the Fermi process in
relativistic shocks including large angle scattering have resulted
in hard electron energy spectra (Stecker, Baring \& Summerlin 2007).

The synchrotron self-Compton (SSC) scattering is not considered
here, as the Compton parameter $Y$ is much less than 1. Without
including Klein-Nishina effects, $Y_{\rm noKN} \simeq (-1+\sqrt{1+4
\epsilon_{rad} \epsilon_e / \epsilon_B})/2$ (Sari \& Esin 2001), the
radiation efficiencies are
$\epsilon_{rad}=(\gamma_c/\gamma_m)^{2-p}$ (only suitable for $p>2$
however) for $\gamma_c>\gamma_m$ and $\epsilon_{rad}=1$ for
$\gamma_c<\gamma_m$ respectively. For the parameters chosen here,
$Y_{\rm noKN} \simeq 7$.
However, for the typical Lorentz factor 
$\gamma_c \simeq 5\times 10^6 t_{\oplus,6}^{7/8}$, and the typical
frequency in the comoving frame
\textbf{$\nu_c' \sim 10^{18} t_{\oplus,6}^{5/8}$ Hz, the main SSC
emission at $\sim 2 \gamma_c^2 \nu_c$ (Sari \& Esin 2001) is deeply
suppressed by the Klein-Nishina effects and the final $Y$ parameter
is much less than 1.} This is also consistent with non-detection by
Fermi and HESS at higher energy bands.

Our model may also be used for the GRBs with long term X-ray flares,
like GRBs 070311 (Vergani \& Guidorzi 2008), and 071118 (Cummings
2007). The off-axis case of our model may also be applied to the
Galactic center, where a gas cloud is ongoing to the central Black
Hole (Gillessen et al. 2012). If this accretion will also produce
episodic jets, the jets will most likely not point to Earth, i.e.,
the off-axis case.

\section*{ACKNOWLEDGMENTS}
We thank the discussion with Xuefeng Wu and Weihua Lei. We thank an
anonymous referee for helpful comments and suggestions. We thank
Kevin MacKeown for a critical reading of the manuscript. This work
made use of data supplied by the UK Swift Science Data Centre at the
University of Leicester. KSC is supported by a GRF grant of Hong
Kong Government under HKU 701013. FYW and YCZ are supported by the
National Natural Science Foundation of China (grant 11103007,
11033002 and U1231101).


\begin{thebibliography}{}
\bibitem[]{}Aliu E., et al. 2011, ApJ, 738, L30
\bibitem[]{}Ayal S., Livio M., \& Piran T., 2000, ApJ, 545, 772
\bibitem[]{}Bhattacharya D., Jun. 2001, Bulletin of the Astronomical Society of India 29, 107
\bibitem[]{}Berger E., et al. 2012, ApJ, 748, 36
\bibitem[]{}Blandford R. D., \& Znajek R. L., 1977, MNRAS, 179, 433
\bibitem[]{}Bloom J. S., et al. 2011, Science, 333, 203
\bibitem[]{}Burrows D. N., et al. 2011, Nature, 476, 421
\bibitem[]{}Cannizzo J. K., Gehrels N., \& Vishniac E. T., 2004, ApJ, 601, 380
\bibitem[]{}Cannizzo J. K., Troja E., \& Lodato G., 2011, ApJ, 742, 32
\bibitem[]{}Covino S., et al., 2010, A\&A, 521, A53
\bibitem[]{}Cummings J. R., et al., 2007, GCN Circular, 7106, 1
\bibitem[]{}Dai Z. G., \& Cheng K. S., 2001, ApJ, 558, L109
\bibitem[]{}De Colle, Fabio., Guillochon J., Naiman J., \& Ramirez-Ruiz E., 2012, ApJ, 760, 103
\bibitem[]{}Evans P. A., et al. 2009, MNRAS, 397, 1177
\bibitem[]{}Gao W. H., 2012, ApJ, 761, 113
\bibitem[]{}Gillessen S., et al., 2012, Nature, 481, 51
\bibitem[]{}Hills J. G. 1975, Nature, 254, 295
\bibitem[]{}Huang Y. F., Cheng K. S., \& Gao T. T., 2006, ApJ, 637, 873
\bibitem[]{}Krolik J. H., \& Piran T., 2011, ApJ, 743, 134
\bibitem[]{}Lacy J. H., Townes C. H., \& Hollenbach, D. J. 1982, ApJ, 262, 120
\bibitem[]{}Lei W. H., \& Zhang B., 2011, ApJ, 740, L27
\bibitem[]{}Lei W. H., Zhang B., \& Gao H., 2013, ApJ, 762, 98
\bibitem[]{}Levan A. J., et al. 2011, Science, 333, 199
\bibitem[]{}Metzger B. D., Giannios D., \& Mimica P. 2012, MNRAS, 420, 3528
\bibitem[]{}Pushkarev A. B., Kovalev Y. Y., Lister M. L. \& Savolainen T., 2009, A\&A, 507, L33
\bibitem[]{}Rees M. J. 1988, Nature, 333, 523
\bibitem[]{}Reis R. C., et al., 2012, Science, 337, 949
\bibitem[]{}Rhoads J. E. 1999, ApJ, 525, 737
\bibitem[]{}Rybicki G. B., \& Lightman A. P., 1979, Radiative Processes in Astrophysics. Wiley, New York
\bibitem[]{}Sari R., \& Esin A. A., 2001, ApJ, 548, 787
\bibitem[]{}Sari R., \& Piran T., 1995, ApJ, 455, L143
\bibitem[]{}Sari R., Piran T., \& Narayan R., 1998, ApJ, 497, L17
\bibitem[]{}Saxton C. J., Soria R., Wu K., \& Kuin N. P. M., 2011, MNRAS, 422, 1625
\bibitem[]{} Shao L., et al. 2011, ApJ, 734, L33
\bibitem[]{}Stecker F. W., Baring M. G., \& Summerlin E. J., 2007, ApJ, 667, L29
\bibitem[]{}Strubbe L. E., \& Quataert E., 2009, MNRAS, 400, 2070
\bibitem[]{}Strubbe L. E., \& Quataert E., 2011, MNRAS, 415, 168
\bibitem[]{}Ulmer A., 1999, ApJ, 514, 180
\bibitem[]{}Vergani S. D., Guidorzi C.,     2008, International Journal of Modern Physics D, 17, 1359
\bibitem[]{}Wang F. Y., \& Cheng K. S., 2012, MNRAS, 421, 908
\bibitem[]{}Wong A. Y. L., Huang Y. F., \& Cheng K. S., 2007, A\&A, 472, 93
\bibitem[]{}Wright E. L., 2006, PASP, 118, 1711
\bibitem[]{}Zauderer B. A., et al. 2011, Nature, 476, 425
\bibitem[]{}Zauderer B. A., Berger E., Margutti R., et al. 2013, ApJ, 767, 152
\bibitem[]{}Zou Y. C., \& Piran T., 2010, MNRAS, 402, 1854
\bibitem[]{}Zou Y. C., Wu X. F., \& Dai Z. G., 2005, MNRAS, 363, 93
\end{thebibliography}
\end{document}